# Landau theory-based estimates for viscosity coefficients of uniaxial and biaxial nematic liquid crystals

Shaikh M. Shamid and David W. Allender[*]

*Department of Physics, Kent State University, Kent, Ohio, USA 44242*

Abstract

Using Landau theory, it is shown that eight phenomenological parameters are needed to describe and distinguish the twelve viscosity coefficients of a biaxial nematic phase, or the five viscosity coefficients of a uniaxial nematic phase. The dependence of the coefficients on the macroscopic uniaxial and biaxial order parameters is established. Since these order parameters are determined by the anisotropies of the dielectric constant, we show that it should be possible to determine values for all eight of the phenomenological parameters of the theory from measurements of the temperature dependence of the five viscosities of a uniaxial phase.

## 1. Introduction

The theory of the viscous hydrodynamic behaviour of uniaxial nematic liquid crystals is fairly well understood. The basic equations have been developed and are given in any standard textbook on liquid crystals, such as de Gennes and Prost [1]. To estimate the expected magnitudes and temperature variations of the nematic viscosity coefficients, an especially useful contribution is the use of Landau theory by Diogo and Martins [2] to derive the dependence of the five viscosity coefficients of a uniaxial nematic phase on the macroscopic order parameter, typically taken to be the anisotropy of the dielectric constant $\Delta\varepsilon$. These theories provide useful

---

guidelines for many applications, including liquid crystal displays. For example, in a display the relaxation of a deformed nematic director is expected to be approximately an exponential decay with a time constant τ, where τ is proportional to the rotational viscosity $\gamma_1$, and inversely proportional to an effective elastic constant which depends on the geometry of the deformation. In fact, one approach to decrease $\tau$ and thereby obtain better performance has been to consider biaxial nematic phases, which in principle have a number of new switching modes and possible geometries. However, the viscous properties of a biaxial nematic have not been developed as much as for uniaxial nematics. This is not surprising, since there have been very few (if any) thermotropic materials that have been shown to have a room temperature biaxial nematic phase. Nevertheless, the visco-elastic properties of a biaxial nematic phase were investigated theoretically in a seminal paper by Saupe [3], following his discovery of a biaxial nematic phase in a lyotropic potassium laurate, decanol and water mixture [4]. He showed that there are twelve independent viscosity coefficients in a biaxial nematic, rather than the well known five of the uniaxial nematic.

Saupe's work has been further developed by Fialkowski [5] and Osipov and Sonnet [6], who derived expressions for the viscosity coefficients in terms of microscopic mean field models based on order parameters describing the averaged orientation of the mesogenic molecules. In this paper a simpler macroscopic model is used. Of course, as described in de Gennes and Prost [1], it is possible to establish a connection between the microscopic and macroscopic models provided the macroscopic property is simply the sum of the effect on individual molecules, as is the case for anisotropic magnetic susceptibility. For the case of dielectric anisotropy, the connection is more complicated and in general not known. The aim of this paper is two fold: (i) to extend the work of Diogo and Martins [2] to include all twelve of Saupe's biaxial nematic viscosities and both uniaxial and biaxial dielectric anisotropies, and (ii) to indicate how the various phenomenological parameters of the Landau theory might be experimentally determined. There is a concern that if the number of adjustable parameters becomes too large, the theory loses its usefulness for application to practical problems.

In the next section, the Landau formalism for the biaxial nematic will be developed and the necessary phenomenological parameters will be introduced. Then in section 3, Saupe's twelve biaxial nematic coefficients will be discussed, including the notation used in naming them, and their relationship to the standard $\alpha_1$, $\alpha_4$, $\alpha_5$, $\gamma_1$, and $\gamma_2$ of a uniaxial nematic. Section 4 gives the results of the calculation: specifically formulas for the dependences of the viscosity coefficients on the macroscopic uniaxial and biaxial order parameters (which are given by the differences among the three principle values of the dielectric constant tensor). Section 5 shows how all the material dependent phenomenological parameters can in principle be determined uniquely by measurements of the $\alpha$'s, $\gamma$'s and dielectric constant anisotropy of the uniaxial nematic phase as functions of temperature.

## 2. Landau Theory of the Dissipation Function in a Biaxial Nematic

As discussed by Saupe [3], the dissipation function $\phi$ of a biaxial nematic is a quadratic function of the time derivatives of the strains and rotations of the local orientational axes in the material. Therefore two tensors are needed to construct the dissipation function: one for the time derivative of the strain, and one for the angular velocity of the alignment axes. The time derivative of the strain is connected with mass flow. It applies to all fluids and is well known to be given by the symmetric part of the velocity gradient tensor: $A_{\alpha\beta} = (\partial_\alpha v_\beta + \partial_\beta v_\alpha)/2$. The angular velocity of alignment axes is not relevant for isotropic fluids, but of course is necessary for liquid crystals and it is useful to describe its representation in more detail.

To describe rotations of the alignment axes, one begins with a local set of mutually orthogonal axes **a**, **b**, and **c**. The liquid crystalline molecules align with respect to these axes, leading to three distinct principle values of the dielectric tensor: $\varepsilon_a > \varepsilon_b > \varepsilon_c$. If the **a, b, c** axes are rotating and the rotation velocities are small, the rotation is given by three scalar rotation speeds, $\Omega_a$, $\Omega_b$ and $\Omega_c$. Here $\Omega_a$ is the angular speed of the b and c axes about the a axis, and similarly for $\Omega_b$ and $\Omega_c$. Therefore $\boldsymbol{\Omega} = \Omega_a \mathbf{a} + \Omega_b \mathbf{b} + \Omega_c \mathbf{c}$, where $\boldsymbol{\Omega}$ is the vector angular velocity. Then the time derivatives are:

$$\partial_t \mathbf{a} = \Omega_c \mathbf{b} - \Omega_b \mathbf{c}$$
$$\partial_t \mathbf{b} = \Omega_a \mathbf{c} - \Omega_c \mathbf{a} \qquad (2.1)$$
$$\partial_t \mathbf{c} = \Omega_b \mathbf{a} - \Omega_a \mathbf{b}$$

The anisotropic part of the velocity gradient tensor is $B_{\alpha\beta} \equiv (\partial_\alpha v_\beta - \partial_\beta v_\alpha)/2$. Defining the vorticity, $\boldsymbol{\omega} = (\nabla \times \mathbf{v})/2$, it is noted that $\omega_\gamma = \varepsilon_{\gamma\alpha\beta} B_{\alpha\beta}$ and vorticity in the fluid is proportional to the angular momentum of the velocity field, so that $\boldsymbol{\omega}$ is the angular velocity of the mass flow. The coupling between rotations of the alignment axes $\boldsymbol{\Omega}$ and the angular velocity of the mass flow $\boldsymbol{\omega}$ is such that there is no decay or dissipation of the angular momentum if $\boldsymbol{\Omega} = \boldsymbol{\omega}$, that is, if the alignment axes rotate together with the principle axes of the velocity gradient tensor.

The last concept needed to construct the tensor for the relative angular velocity of the alignment axes is the Landau-de Gennes tensor for a biaxial nematic phase, $Q_{\alpha\beta}$ [7]. Different researchers have used a variety of representations for this tensor, but in the present work a convenient choice is:

$$Q_{\alpha\beta} = (\varepsilon_a - \varepsilon_{ave})a_\alpha a_\beta + (\varepsilon_b - \varepsilon_{ave})b_\alpha b_\beta + (\varepsilon_c - \varepsilon_{ave})c_\alpha c_\beta \qquad (2.2)$$

where $\varepsilon_{ave} = (\varepsilon_a + \varepsilon_b + \varepsilon_c)/3$. For convenience, it can be assumed that $(\varepsilon_a - \varepsilon_b) > (\varepsilon_b - \varepsilon_c)$, so the standard definitions $q \equiv (\varepsilon_a - \varepsilon_{ave}) > 0$ and $p \equiv (\varepsilon_b - \varepsilon_c) > 0$ can be made, giving

$$Q_{\alpha\beta} = q\, a_\alpha a_\beta + \left(\frac{-q+p}{2}\right)b_\alpha b_\beta + \left(\frac{-q-p}{2}\right)c_\alpha c_\beta \qquad (2.3)$$

Here $q$ and $p$ are the macroscopic order parameters, clearly based on the anisotropy of the dielectric tensor, $q \neq 0$ and $p=0$ for the uniaxial nematic, but both $q$ and $p$ are non-zero for the biaxial nematic. The symbol $q$ rather than S has deliberately been chosen for the uniaxial phase in order to emphasize that it is a macroscopic order parameter. The symbol S is standardly used to denote the Maier-Saupe order parameter describing the average over all molecules of the alignment of the molecular long axis with respect to the optic axis. Under some assumptions, one can show that $q$ is proportional to S [1, 7]. It is now straightforward to take the time derivative of $Q_{\alpha\beta}$ using eq. (2.1) and (2.3). Taking care to replace $\boldsymbol{\Omega}$ with $\boldsymbol{\Omega} - \boldsymbol{\omega}$ since dissipation depends only on this difference, one finally obtains the expression for the tensor, $W_{\alpha\beta}$ which represents the time derivative of the angles specifying the relative rotations of the orientation axes:

$$W_{\alpha\beta} = \left(\frac{3q-p}{2}\right)(\Omega_c - \omega_c)(a_\alpha b_\beta + b_\alpha a_\beta) - \left(\frac{3q+p}{2}\right)(\Omega_b - \omega_b)(a_\alpha c_\beta + c_\alpha a_\beta)$$

$$+ p(\Omega_a - \omega_a)(b_\alpha c_\beta + c_\alpha b_\beta) \qquad (2.4)$$

Saupe [3] expressed the vector $\boldsymbol{\Omega} - \boldsymbol{\omega}$ differently by introducing a vector **N** defined as follows:

$$\mathbf{N} = (\Omega_a - \omega_a)\mathbf{c} + (\Omega_b - \omega_b)\mathbf{a} + (\Omega_c - \omega_c)\mathbf{b} \qquad (2.5)$$

where $(\Omega_a - \omega_a) = (\partial_t \mathbf{b} - \boldsymbol{\omega} \times \mathbf{b}) \cdot \mathbf{c}$, along with cyclic permutation of symbols for the **a** and **b** components of **N**. Thus

$$W_{\alpha\beta} = \left(\frac{3q-p}{2}\right)(\mathbf{N}\cdot\mathbf{b})(a_\alpha b_\beta + b_\alpha a_\beta) - \left(\frac{3q+p}{2}\right)(\mathbf{N}\cdot\mathbf{a})(a_\alpha c_\beta + c_\alpha a_\beta)$$

$$+ p(\mathbf{N}\cdot\mathbf{c})(b_\alpha c_\beta + c_\alpha b_\beta) \qquad (2.6)$$

This explanation of $W_{\alpha\beta}$ has implicitly assumed that $q$ and $p$ do not depend on either space or time. They are simply temperature dependent macroscopic order parameters. This assumption is expected to be valid when defects are not present and when surfaces provide only a direction for orientation of alignment axes but do not modify significantly the degree of molecular alignment.

As a last approximation, the fluid can be treated as incompressible, which makes the $A_{\alpha\beta}$ tensor traceless. Thus both the $A_{\alpha\beta}$ and $W_{\alpha\beta}$ tensors are symmetric and traceless. According to the principles of Landau theory, the dissipation function $\phi$ can now be obtained by forming all scalar contractions of the $A_{\alpha\beta}$ and $W_{\alpha\beta}$ tensors, out to quadratic order in A and W, i.e. schematically AA, AW and WW. Note that AA is independent of $q$ and $p$, while AW is linear in $q$ and $p$, and WW is quadratic in $q$ and $p$. Additional contractions including factors of $Q_{\alpha\beta}$ i.e. of the form $Q$AA, $QQ$AA, $Q$AW, must be considered in order to consistently get all terms out to quadratic order in $q$ and $p$. Of course contractions with additional factors of $Q$ could be included, but it is assumed that only the minimum number of terms necessary to produce the observed phenomena will be kept in order to keep the number of phenomenological parameters to a minimum. As will be seen, quadratic order in $q$ and $p$ is necessary and sufficient to produce independent values of all twelve viscosity coefficients.

Out to quadratic order in $q$ and $p$, the number of possible contractions is nine. They are:

$$\begin{aligned}
I_1 &= A_{\alpha\beta} A_{\beta\alpha} \\
I_2 &= Q_{\alpha\beta} A_{\beta\mu} A_{\mu\alpha} \\
I_3 &= \left(Q_{\alpha\beta} Q_{\beta\alpha}\right)\left(A_{\mu\nu} A_{\nu\mu}\right) \\
I_4 &= \left(Q_{\alpha\beta} A_{\beta\alpha}\right)\left(Q_{\mu\nu} A_{\nu\mu}\right) \\
I_5 &= Q_{\alpha\beta} Q_{\beta\mu} A_{\mu\nu} A_{\nu\alpha} \\
I_6 &= W_{\alpha\beta} W_{\beta\alpha} \\
I_7 &= A_{\alpha\beta} W_{\beta\alpha} \\
I_8 &= Q_{\alpha\beta} A_{\beta\mu} W_{\mu\alpha} \\
I_9 &= Q_{\alpha\beta} A_{\beta\mu} Q_{\mu\nu} A_{\nu\alpha}
\end{aligned} \qquad (2.7)$$

The standard convention of an implicit sum over a repeated index is used. It will be shown in section 4 that upon evaluating the nine contractions, $I_9$ is not independent, but is in fact expressible as the linear combination

$$2I_9 = I_3 + 2I_4 - 4I_5 . \qquad (2.8)$$

Therefore only eight contractions are independent. The dissipation function is then given by

$$2\phi = J_1 I_1 + J_2 I_2 + \ldots\ldots + J_8 I_8 \qquad (2.9)$$

where the $J_i$'s are phenomenological constants associated with the eight contractions.

## 3. Viscosity Coefficients of a Biaxial Nematic

After a lengthy derivation, Saupe in reference [3], gave an expression for the twelve viscosity coefficients of an incompressible biaxial nematic in his eq. (42), which is an equation for twice the dissipation function, $2\phi$. There are twelve terms in $2\phi$ with each term being of the form of a viscosity coefficient multiplied by scalar terms that come from contractions that are quadratic in the elements of the A and W tensors. Technically, eq. (42) allowed for compressibility and contains another three terms for a total of fifteen; but as noted in [3], those extra three terms vanish for an incompressible fluid. In an attempt to give some physical intuition to the meaning of the different viscosities, the conclusions reached in [3] will be reiterated here. Specifically, each of the twelve terms will be examined. Three of the viscosities can be found by measuring the time constants for exponential decay of twist deformations about the **a**, **b**, and **c** axes, respectively, assuming that **v** is either zero, or can be suppressed. The remaining nine viscosities can be measured by performing shear experiments on aligned samples having parallel plate substrates with the top surface moving relative to the bottom surface. Nine possible geometries are needed. For example let **a** be parallel to the substrates but normal to the velocity of the moving surface. Perform the experiment with three different orientations of **b**: (1) parallel to the velocity of the moving surface, (2) normal to the substrate, and (3) at some angle in between, say $\pi/4$. The first orientation defines a viscosity $\eta_{bc}$, the second a viscosity $\eta_{cb}$ and the third an effective viscosity $\eta_{eff}$ parameterized by $\alpha_{bc}$ where $\eta_{eff} = \left(-\alpha_{bc}/2 + \eta_{bc} + \eta_{cb}\right)/2$ (Ref. 3 gives expressions for an arbitrary angle for **b**, not just $\pi/4$, which is assumed here). The indices on the $\eta_{bc}$ mean that **v** is parallel to **b** and the gradient of **v** is parallel to **c**. Therefore for **a** oriented as stated, three viscosities, $\eta_{bc}$, $\eta_{cb}$ and $\alpha_{bc}$ can be obtained. By permuting **a** to **b** and **c**, the other six viscosities can be found. Thus, in principle, all twelve coefficients are obtainable if the biaxial phase can be oriented as desired. The only remaining identification that must be made is to show how the $\eta$'s and $\alpha$'s are related to the coefficients in the expression for the dissipation function 2ϕ.

To simplify expressions, the indices of tensor contractions will be omitted, using the notation: $x_\alpha T_{\alpha\beta} y_\beta \equiv \mathbf{x T y} = \mathbf{y T x}$, where **x** and **y** are vectors, and **T** is a symmetric rank two tensor.

Three of the twelve terms in $2\phi$ are:

$$\gamma_{aa}\{(\Omega-\omega)\cdot\mathbf{a}\}^2$$
$$\gamma_{bb}\{(\Omega-\omega)\cdot\mathbf{b}\}^2 \qquad (3.1)$$
$$\gamma_{cc}\{(\Omega-\omega)\cdot\mathbf{c}\}^2$$

Clearly, the factor multiplying $\gamma_{aa}$ is independent of any shear and depends only on $\Omega$ and $\omega$, so $\gamma_{aa}$ is simply the viscosity arising due to the rotation of the **b** and **c** axes about the **a** axis, relative to the **a** component of mass flow angular velocity $\omega$. Similarly, $\gamma_{bb}$ and $\gamma_{cc}$ are viscosities for rotations of alignment axes about **b** and **c** respectively. These correspond precisely to the relaxation experiments mentioned above.

The remaining nine terms in $2\phi$ are:

$$2\gamma_{abc}\{(\Omega-\omega)\cdot\mathbf{a}\}(\mathbf{bAc})$$
$$2\gamma_{bca}\{(\Omega-\omega)\cdot\mathbf{b}\}(\mathbf{cAa})$$
$$2\gamma_{cab}\{(\Omega-\omega)\cdot\mathbf{c}\}(\mathbf{aAb})$$
$$\sigma_{ab}(\mathbf{aAa})(\mathbf{bAb})\equiv(2\eta_{aabb}-\eta_{aaaa}-\eta_{bbbb})(\mathbf{aAa})(\mathbf{bAb})$$
$$\sigma_{bc}(\mathbf{bAb})(\mathbf{cAc})\equiv(2\eta_{bbcc}-\eta_{bbbb}-\eta_{cccc})(\mathbf{bAb})(\mathbf{cAc}) \qquad (3.2)$$
$$\sigma_{ca}(\mathbf{cAc})(\mathbf{aAa})\equiv(2\eta_{ccaa}-\eta_{cccc}-\eta_{aaaa})(\mathbf{cAc})(\mathbf{aAa})$$
$$4\eta_{abab}(\mathbf{aAb})^2$$
$$4\eta_{bcbc}(\mathbf{bAc})^2$$
$$4\eta_{caca}(\mathbf{cAa})^2$$

where the γ and η notation was defined by Saupe [3]. Saupe showed that the connection between the viscosities that would be measured in the shear experiments discussed above and the coefficients in the $2\phi$ terms are:

$$\eta_{ab} = \eta_{abab} + \frac{\gamma_{cc}}{4} + \frac{\gamma_{cab}}{2}$$

$$\eta_{ba} = \eta_{abab} + \frac{\gamma_{cc}}{4} - \frac{\gamma_{cab}}{2} \qquad (3.3)$$

$$\alpha_{ab} = -\sigma_{ab} - 4\eta_{abab}$$

Thus $\eta_{ab}$, $\eta_{ba}$ and $\alpha_{ab}$ values yield numbers for $\eta_{abab}$, $\gamma_{cab}$ and $\sigma_{ab}$. Recall that $\gamma_{cc}$ was obtained separately in the rotational relaxation experiment. Cyclic permutation of **a**, **b**, and **c** gives the other eight viscosities.

## 4. Order Parameter Dependences of the Viscosities

It is tedious but straightforward to evaluate each of the nine scalar invariants introduced in section 2, expressing them in a way compatible with the dissipation function $2\phi$ discussed in section 3. For example, $I_1$ is evaluated as follows:

$$I_1 = (\mathbf{aAa})^2 + (\mathbf{bAb})^2 + (\mathbf{cAc})^2 + 2(\mathbf{aAb})^2 + 2(\mathbf{bAc})^2 + 2(\mathbf{cAa})^2$$

Then noting that the incompressible fluid approximation gives $\nabla \cdot \mathbf{v} = (\mathbf{aAa}) + (\mathbf{bAb}) + (\mathbf{cAc}) = 0$, the relation $(\mathbf{aAa})^2 = -(\mathbf{aAa})(\mathbf{bAb}) - (\mathbf{cAc})(\mathbf{aAa})$ is obtained, yielding

$$I_1 = 2\{(\mathbf{aAb})^2 + (\mathbf{bAc})^2 + (\mathbf{cAa})^2 - (\mathbf{aAa})(\mathbf{bAb}) - (\mathbf{bAb})(\mathbf{cAc}) - (\mathbf{cAc})(\mathbf{aAa})\}$$

Similarly, the other eight contractions $I_2$ to $I_9$ may be evaluated. The results are:

$$I_1 = 2\left\{(\mathbf{aAb})^2 + (\mathbf{bAc})^2 + (\mathbf{cAa})^2 - (\mathbf{aAa})(\mathbf{bAb}) - (\mathbf{bAb})(\mathbf{cAc}) - (\mathbf{cAc})(\mathbf{aAa})\right\}$$

$$I_2 = \frac{q+p}{2}\left\{(\mathbf{aAb})^2 - (\mathbf{aAa})(\mathbf{bAb})\right\} + \frac{q-p}{2}\left\{(\mathbf{cAa})^2 - (\mathbf{aAa})(\mathbf{cAc})\right\} - q\left\{(\mathbf{bAc})^2 - (\mathbf{bAb})(\mathbf{cAc})\right\}$$

$$I_3 = (3q^2 + p^2)\left\{(\mathbf{aAb})^2 + (\mathbf{bAc})^2 + (\mathbf{cAa})^2 - (\mathbf{aAa})(\mathbf{bAb}) - (\mathbf{bAb})(\mathbf{cAc}) - (\mathbf{cAc})(\mathbf{aAa})\right\}$$

$$I_4 = -\frac{1}{4}\left\{(3q-p)^2 (\mathbf{aAa})(\mathbf{bAb}) + (3q+p)^2 (\mathbf{aAa})(\mathbf{cAc}) + 4p^2 (\mathbf{bAb})(\mathbf{cAc})\right\}$$

$$I_5 = \frac{1}{4}\left[(5q^2 - 2qp + p^2)\left\{(\mathbf{aAb})^2 - (\mathbf{aAa})(\mathbf{bAb})\right\} + (5q^2 + 2qp + p^2)\left\{(\mathbf{cAa})^2 - (\mathbf{aAa})(\mathbf{cAc})\right\}\right]$$
$$+ \frac{1}{2}(q^2 + p^2)\left\{(\mathbf{bAc})^2 - (\mathbf{bAb})(\mathbf{cAc})\right\}$$

$$I_6 = \frac{1}{2}\left[(3q-p)^2 \left\{(\mathbf{\Omega}-\mathbf{\omega})\cdot\mathbf{c}\right\}^2 + (3q+p)^2 \left\{(\mathbf{\Omega}-\mathbf{\omega})\cdot\mathbf{b}\right\}^2 + 4p^2 \left\{(\mathbf{\Omega}-\mathbf{\omega})\cdot\mathbf{a}\right\}^2\right]$$

$$I_7 = (3q-p)\left\{(\mathbf{\Omega}-\mathbf{\omega})\cdot\mathbf{c}\right\}(\mathbf{aAb}) - (3q+p)\left\{(\mathbf{\Omega}-\mathbf{\omega})\cdot\mathbf{b}\right\}(\mathbf{cAa}) + 2p\left\{(\mathbf{\Omega}-\mathbf{\omega})\cdot\mathbf{a}\right\}(\mathbf{bAc})$$

$$I_8 = \frac{1}{4}(3q-p)(q+p)\left\{(\mathbf{\Omega}-\mathbf{\omega})\cdot\mathbf{c}\right\}(\mathbf{aAb})$$
$$- \frac{1}{4}\left[(3q+p)(q-p)\left\{(\mathbf{\Omega}-\mathbf{\omega})\cdot\mathbf{b}\right\}(\mathbf{cAa}) + 4qp\left\{(\mathbf{\Omega}-\mathbf{\omega})\cdot\mathbf{a}\right\}(\mathbf{bAc})\right]$$

$$I_9 = -q(q-p)(\mathbf{aAb})^2 - q(q+p)(\mathbf{cAa})^2 + \frac{1}{2}(q^2 - p^2)(\mathbf{bAc})^2$$
$$- \frac{1}{4}\left[(5q^2 - 2qp + p^2)(\mathbf{aAa})(\mathbf{bAb}) + (5q^2 + 2qp + p^2)(\mathbf{aAa})(\mathbf{cAc}) + 2(q^2 + p^2)(\mathbf{bAb})(\mathbf{cAc})\right]$$

(4.1)

Inspection reveals that $I_3 + 2I_4 - 4I_5 - 2I_9 = 0$. Note that each of the terms appearing in the $I_1$ to $I_9$ contractions is the same as the factors that multiply the viscosities in the expression for $2\phi$ in eq. (3.2).

Lastly, inserting these expressions for the $I$'s into Eq. (2.9) and comparing to the form of $2\phi$ from [3], the desired expressions for the viscosity coefficients are revealed:

$$\sigma_{ab} = -2J_1 - \frac{J_2}{2}(q+p) - J_3(3q^2+p^2) - \frac{J_4}{4}(3q-p)^2 - \frac{J_5}{4}(5q^2-2qp+p^2)$$

$$\sigma_{bc} = -2J_1 + J_2 q - J_3(3q^2+p^2) - J_4 p^2 - \frac{J_5}{2}(q^2+p^2)$$

$$\sigma_{ca} = -2J_1 - \frac{J_2}{2}(q-p) - J_3(3q^2+p^2) - \frac{J_4}{4}(3q+p)^2 - \frac{J_5}{4}(5q^2+2qp+p^2)$$

$$4\eta_{abab} = 2J_1 + \frac{J_2}{2}(q+p) + J_3(3q^2+p^2) + \frac{J_5}{4}(5q^2-2qp+p^2)$$

$$4\eta_{bcbc} = 2J_1 - J_2 q + J_3(3q^2+p^2) + \frac{J_5}{2}(q^2+p^2)$$

$$4\eta_{caca} = 2J_1 + \frac{J_2}{2}(q-p) + J_3(3q^2+p^2) + \frac{J_5}{4}(5q^2+2qp+p^2) \qquad (4.2)$$

$$\gamma_{aa} = 2J_6 p^2$$

$$\gamma_{bb} = \frac{J_6}{2}(3q+p)^2$$

$$\gamma_{cc} = \frac{J_6}{2}(3q-p)^2$$

$$2\gamma_{abc} = 2J_7 p - J_8 qp$$

$$2\gamma_{bca} = -J_7(3q+p) - \frac{J_8}{4}(3q+p)(q-p)$$

$$2\gamma_{cab} = J_7(3q-p) + \frac{J_8}{4}(3q-p)(q+p)$$

There are several features that should be noticed. The three η's and the three σ's all have a term independent of $q$ and $p$, and therefore combine to give the viscosity of the isotropic phase. Each of them also has contributions linear in $q$ and $p$, and quadratic in $q$ and $p$. All six γ's go to zero in the isotropic phase as expected because alignment axes do not exist in the isotropic phase. The $\gamma_{aa}$ rotational viscosity also vanishes in the uniaxial nematic phase, as does $\gamma_{abc}$. Furthermore as $p$ goes to zero, $\sigma_{ab} = \sigma_{ca}$, $4\eta_{bcbc} = -\sigma_{bc}$, $\eta_{abab} = \eta_{caca}$, $\gamma_{bb} = \gamma_{cc}$ and $\gamma_{cab} = -\gamma_{bca}$. This leaves five independent viscosities for the uniaxial nematic phase, as expected. The result is:

$$\alpha_1 = -(\sigma_{ab} + 4\eta_{abab}) = \frac{9}{4} J_4 q^2$$

$$\alpha_4 = 2\eta_{bcbc} = J_1 - \frac{1}{2} J_2 q + \left(\frac{3}{2} J_3 + \frac{1}{4} J_5\right) q^2$$

$$\alpha_5 + \alpha_6 = 4(\eta_{abab} - \eta_{bcbc}) = \frac{3}{2} J_2 q + \frac{3}{4} J_5 q^2 \quad (4.3)$$

$$\gamma_1 = \gamma_{bb} = \frac{9}{2} J_6 q^2$$

$$\gamma_2 = \gamma_{cab} = \frac{3}{2} J_7 q + \frac{3}{8} J_8 q^2$$

where the first equality in each line is given by Eq. (55) of [3] and the second equality of each line is our findings.

## 5. Determination of the phenomenological parameters

A method to uniquely extract values of the eight $J$'s for a given material can be logically developed based on an extension of the method used by Berreman and Meiboom [8] to determine the Ginzberg-Landau parameters occurring in the expansion for the elastic free energy. All eight constants can be found by fitting to data taken in the uniaxial nematic and isotropic phases, leaving no adjustable parameters for data in the biaxial nematic phase. The relevant information is completely contained in eq. (4.3). $J_1$ can be found by measuring the shear viscosity in the isotropic phase where $q = 0$. The measurements needed are: $q$, $\alpha_1$, $\alpha_4$, $\alpha_5 + \alpha_6$, $\gamma_1$ and $\gamma_2$, each as a function of temperature in the uniaxial nematic phase. This of course would be quite difficult to do. Nonetheless, if the data is available, then Eq. (4.3) can be manipulated to give:

$$\frac{\alpha_1}{q} = \frac{9}{4}J_4 q$$

$$\frac{\alpha_4 - J_1}{q} = -\frac{1}{2}J_2 + \left(\frac{3}{2}J_3 + \frac{1}{4}J_5\right)q$$

$$\frac{\alpha_5 + \alpha_6}{q} = \frac{3}{2}J_2 + \frac{3}{4}J_5 q \qquad (4.4)$$

$$\frac{\gamma_1}{q} = \frac{9}{2}J_6 q$$

$$\frac{\gamma_2}{q} = \frac{3}{2}J_7 + \frac{3}{8}J_8 q$$

From intercepts and slopes of the linear relationships, all eight parameters can be obtained. There is even a redundancy, since two of the intercepts are given by $J_2$. It is not realistic to expect to have sufficient data to actually carry this procedure out, but the intent is simply to demonstrate the consistency of the approach.

## 6. Conclusions

The Landau theory for the dissipation function of a biaxial nematic liquid crystal has been developed. Expressions out to quadratic order in $q$ and $p$, the uniaxial and biaxial macroscopic order parameters, have been obtained for all twelve viscosity coefficients. This information can be coupled with similar knowledge of the $q$ and $p$ dependence of the elastic constants [9] to estimate the decay times of a very large set of possible geometries of LCD cells using a biaxial nematic liquid crystal.

**Acknowledgments**

DWA and SMS thank Samsung Electronics Company for support to carry out this research.